\documentclass[prl,twocolumn,showpacs,preprintnumbers,amsmath,amssymb]{revtex4}

\usepackage{graphicx}
\usepackage{dcolumn}
\usepackage{bm}

\begin{document}
\preprint{}
\pagenumbering{arabic}

\title{Stacking Sequence Dependence of Graphene Layers on SiC($000\overline{1}$) - Experimental and Theoretical Investigation }
\author{Jolanta Borysiuk$^{1, 2}$, Jakub So\l{}tys $^3$, and Jacek Piechota $^3$}
\affiliation{%
$^1$ Institute of Physics, Polish Academy of Sciences, Al. Lotnik\'{o}w 32/46, 02-668, Warsaw, Poland \\
$^2$ Institute of Experimental Physics, Warsaw University, Ho\.{z}a 69, 00-681 Warsaw, Poland \\
$^3$ Interdisciplinary Centre for Mathematical and Computational Modelling, University of Warsaw, Pawi\'nskiego 5a, 02-106 Warsaw, Poland  
}%

\date{\today}

\begin{abstract}
Different stacking sequences of graphene are investigated a using combination of experimental and theoretical methods. The high-resolution transmission electron microscopy (HRTEM) of the stacking sequence
of several layers of graphene, formed on the C-terminated 4H-SiC($000\overline{1}$) surface, 
was used to determine the stacking sequence and the interlayer distances. These data prove that 
the three metastable configurations exist: ABAB, AAAA, ABCA. In accordance to these findings, 
those three cases were considered theoretically, using Density Functional Theory calculations 
comparing graphene sheets, freestanding and positioned on the SiC($000\overline{1}$) substrate. 
The total energies were calculated, the most stable structure was identified and the electronic 
band structure was obtained. The four graphene layer electron band structure depends crucially 
on the stacking: for the ABAB and ABCA stacking, the bands, close to the K point, are characterized 
by the hyperbolic dispersion relation while the AA stacking the dispersion in this region is linear, 
similar to that of a single graphene layer. It was also shown that the linear dispersion relation 
is preserved in the presence of the SiC substrate, and also for different distances between adjacent carbon layers. 
\end{abstract}

\maketitle

Graphene is one of the most extensively investigated perspective materials at the moment. 
Its unique electronic properties, promising very important applications in electronics, 
have attracted interests of many scientific groups. The massless fermion dispersion, observed in 
the vicinity of the K point in the energy range close to the Fermi surface, and the related, 
experimentally determined the $\sqrt{B}$ dependence of the Landau levels, was reported from 
investigations of a single, freestanding graphene layer \cite{Novoselov2,Morozov1,Zhang2}.
In order to implement graphen in the electronics industry, a method 
for the synthesis of reproducible, good quality graphene layers, stabilized on a solid substrate, 
has yet to be developed. The most natural one is the deposition of the epitaxial graphene 
on the SiC substrate, either on the Si- or C-terminated principal faces of the SiC crystal which is 
currently considered as one of the most promising method to attain this goal 
\cite{Hass1,Hass2,Borysiuk1,Varchon1,Varchon2}.
A recent work \cite{Lin} shows that the method is potentially able to furnish good quality devices. 

Graphene multilayer structures have been studied both experimentally and theoretically. 
From these studies, it is recognized that the multilayer graphene electronic properties are 
strongly dependent on the stacking sequence. 
Both the Bernal (AB) and rhombohedral (ABC) stacking sequences are described theoretically 
\cite{Latil1,Latil2}  and experimentally \cite{Mak,Norimatsu}. It was also shown that 
for rhombohedral(ABC) stacking the  
band gap can be induced by an external electric field \cite{Aoki}.   
The description of the AA structure was often overlooked because it is energetically 
unfavourable \cite{Charlier}.
\begin{figure*}
\includegraphics[width=0.65 \textwidth]{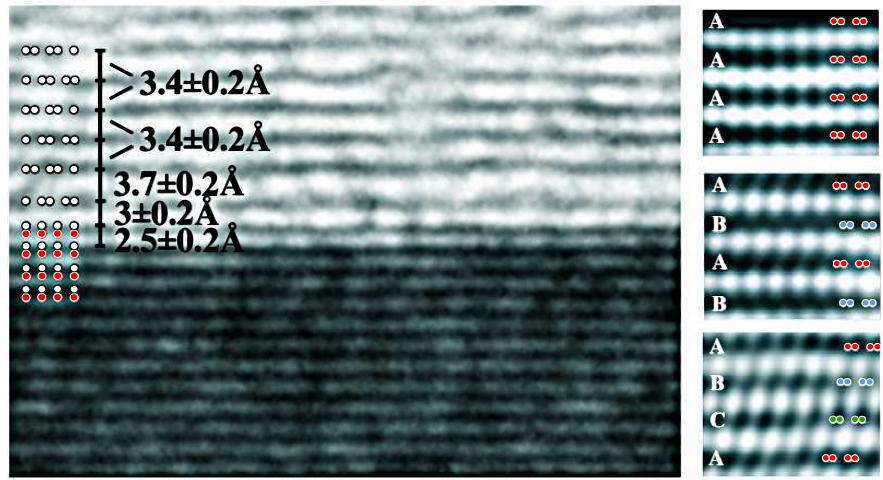}
\caption{\label{fig:stacks} 
(Colour online) (left) A cross-sectional HRTEM image of graphene layers  , grown on the C-terminated 
face of 4H-SiC. The first graphene layer is separated by 3.0\AA \  from the SiC surface, 
next - by 3.7\AA \ from the previous one, and the subsequent layers are separated equally by 3.4\AA \
. (right) Experimental HRTEM filtered images showing different stacking sequences of graphene layers. 
The positions of the carbon atoms columns are marked by circles. }
\end{figure*}
However, very recently, Norimatsu and Kusunoki \cite{Norimatsu} have observed the existence of the AA 
stacking sequence around the step on the Si-face.  We have undertaken systematic investigations of the 
four layer graphene synthesized on the C-terminated 4H-SiC($000\overline{1}$) surface, such 
as the one which has been investigated recently\cite{Borysiuk2}. By an optical inspection, the three 
different regions were identified, which are then subjected to the HRTEM scrutiny. The results 
are presented in Fig. \ref{fig:stacks}. The three different stacking sequences are identified, 
showing the existence of the AA stacking sequence as well as the Bernal (AB) and rhombohedral (ABC) 
sequences, independent of the presence of the steps. Thus these three structures could be 
synthesized on the C-terminated 4H-SiC($000\overline{1}$) surface proving that these structures 
are metastable, and could be obtained by the appropriate selection of the thermodynamic parameters 
of SiC annealing \cite{Borysiuk1}. These structures could persist for macroscopically long times. 

The HRTEM investigations were also used to determine the distances between the graphene sheets. 
The example of such investigation, shown in Fig. \ref{fig:stacks}, presents the determination 
of the carbon interlayer distances. The high precision measurements prove that the carbon bottommost 
layer is separated by a relatively large distance of 3.0 \AA \ from the SiC surface, which is close to 
the interlayer spacing in graphite, the situation which is a standard for graphene on the C-terminated 
face \cite{Borysiuk2}. Therefore these graphene layers are not strongly bound to the underlying SiC 
surface, in contrast to the layers grown on the opposite, Si-terminated surface \cite{Borysiuk1}. 
The following carbon sheet is separated by a distance of 3.7\AA, while the remaining one by 3.4\AA, 
i.e. essentially equal to the carbon sheet separation in graphite. These experimentally determined 
graphene interlayer distances were used in the DFT ab initio calculations described below. 
In the present work, we employ \textit{ab initio} the density functional theory the VASP 
\cite{Kresse1993, Kresse1996, Kresse1996a, Kresse1999} code to investigate graphene-SiC interface. 
We have used the projector augmented wave (PAW) approach \cite{Blochl} in its variant available 
in the VASP package \cite{Kresse1999}. The local spin density approximation (LSDA) was applied for 
the exchange-correlation functional. The plane wave cutoff energy was set to 500 eV. The Monkhorst-Pack 
k-point mesh was set to $7\times7\times1$. The 4H-SiC(0001) supercell was constructed using 8 bilayers 
of Si-C. 
Four graphene layers were located at the top of the SiC($000\overline{1}$) surface, at the 
separation determined experimentally (see Fig. \ref{fig:stacks}). Since the graphene interlayer 
distance results from the van der Waals interaction, which, as a rule, could not be obtained from DFT 
calculations properly, the combined experimental-theoretical approach is the only possible way to obtain 
the properties of multilayer graphene with high precision. In the present calculations, the slab 
replicas are separated by the space of about 17 \AA. An elastic adjustment was performed at the interface 
due to the lattice mismatch between SiC and graphite such that the two top SiC layers and the graphene 
layer was relaxed in the plane. The conjugate gradient algorithm was used in the relaxation of the 
atomic positions. The model, $ \sqrt{3} \times \sqrt{3} R 30 ^\circ - SiC$ unit cell with a fitted 
graphene layer (GL) \cite{Mattaush,Li2,Forbeaux}, was used. 
\begin{figure*}
\includegraphics[width=0.78\textwidth]{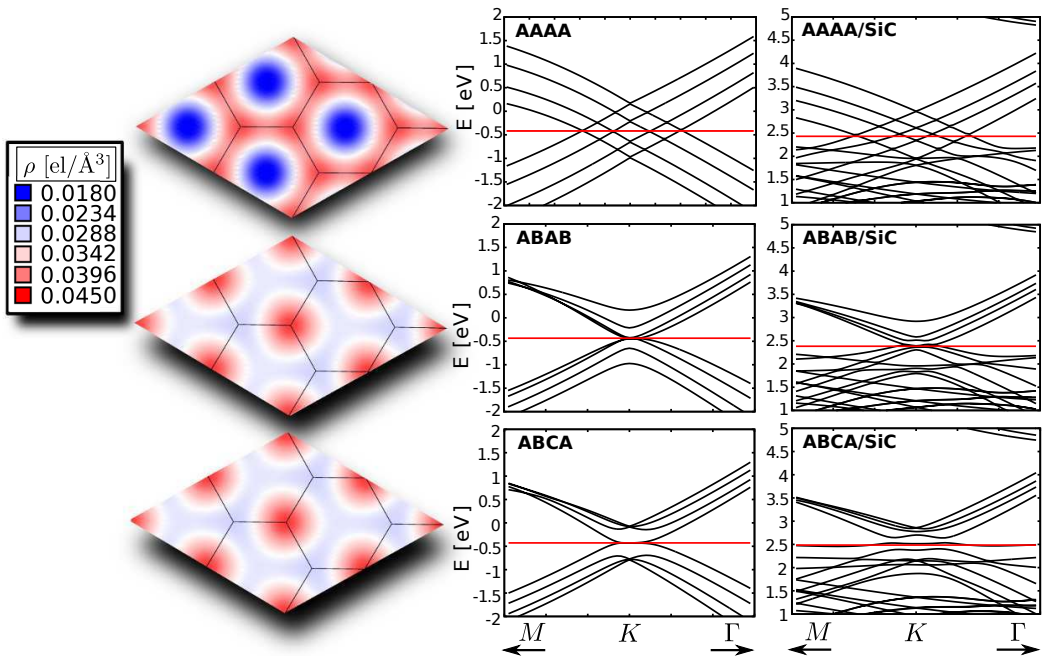}
\caption{\label{fig:bands} (Colour online) Left - the spatial distribution of electron density, 
right - band structures of four graphene layer system in different stacking sequences:  
AAAA (top), ABAB (middle), ABCA (bottom); the left column - freestanding graphene, the right column - 
graphene deposited on the SiC($000\overline{1}$) substrate. }
\end{figure*}
The direct DFT calculation indicate that for freestanding graphene layers, Bernal (ABAB) stacking 
sequence is the most stable, having the total energy equal to \textit{-80.824 eV} which is lower than 
the energy of both the rhombohedral (ABCA)  \textit{-80.836 eV} and (AAAA) \textit{-80.782 eV} stackings, 
which is consistent with the earlier works \cite{Aoki}. 
The electron energy bands are analyzed using a comparison between the results obtained for 
four graphene layers, freestanding and deposited on the 4H-SiC($000\overline{1}$) surface as presented 
in Fig. \ref{fig:bands}. The three stacking configurations, AAAA, ABAB and ABCA exhibit the  miscellaneous 
electronic structure due to the different symmetry.  For the case of the AAAA configuration, the 
Dirac-type dispersion relation is observed. This is in agreement with a simple, tight binding 
argument presented recently by Gonzalez et al \cite{Gonzalez}. As expected, the band structure is 
different for both the Bernal (ABAB) and rhombohedral (ABCA) structures. In the case of the Bernal (ABAB) 
structure, in accordance to the earlier results, both the conduction and valence bands are 
hyperbolic \cite{Mak}. The zero bandgap structure has the Fermi energy located at the common 
maximum of the valence and  the minimum of the conduction bands. A slightly different picture is obtained 
for the freestanding rhombohedral (ABCA) sequence. The hyperbolic dependence, similar to that observed 
for the Bernal stacking is preserved. The two bands constitute a zero bandgap structure at the K point. 
Other bands are collapsed at the K point in the so-called "wizard-hat" shape \cite{Mak}. 
Due to the lower crystallographic symmetry of the rhombohedral structure, they have the energy 
minimum shifted away from the K point (see Fig. \ref{fig:bands} ). 

The influence of the 4H-SiC substrate is different in these three cases. In the case of the  (AAAA) 
stacking, the presence of the SiC substrate amounts to mere addition of the additional energy 
bands, of their energies located in the valence band energies of graphene. In the case of Bernal 
stacking, the influence of the substrate leads to a slight modification of the band structure, 
close to the K point. Thus the epitaxial graphene on the SiC($000\overline{1}$) surface is 
characterized by the essentially identical band structure for both (AAAA) and  Bernal (ABAB), 
in respect to the freestanding graphene in these configurations, that indicates 
a relatively weak coupling of the graphene with the SiC surface 
as suggested by the relatively large distance measured by the HRTEM (see Fig. \ref{fig:stacks}). 
In contrast to that, in the case of rhombohedral stacking, the structure is most affected, 
leading to the opening of the gap, and the creation of the double minimum in the conduction band. 
The second conduction band is concave and it is not on the Fermi level (see. Fig. \ref{fig:bands}). 
The first valence band which is curved up at the K point, is associated with the SiC substrate influence. 


In order to explain why the symmetry of the systems so strongly affects the electronic structure, 
we have calculated electron densities between the first two freestanding graphene layers in AAAA, ABAB, ABCA  
stacking sequence Fig. \ref{fig:bands} (left). In order to compare the electron density distribution 
in these three cases, we have used a consistent scale. For the AAAA stacking sequence, the areas of the increased 
electron density (red pattern) are arranged in the honeycomb
pattern. This electron distribution pattern is similar to the isolated one graphene layer.
Indeed, such a "highway" for electrons results in linear dispersion bands.
The situation is quite different for graphene layers arranged into ABAB and ABCA sequences. 
Electron distributions for these two cases are similar and they form isolated 
islands of the increased electron density (red pattern). This different electron configuration gives 
hyperbolic HOMO and LUMO bands dispersion (see Fig \ref{fig:bands} (left) ).

\begin{figure}[t]
\includegraphics[width=0.405 \textwidth]{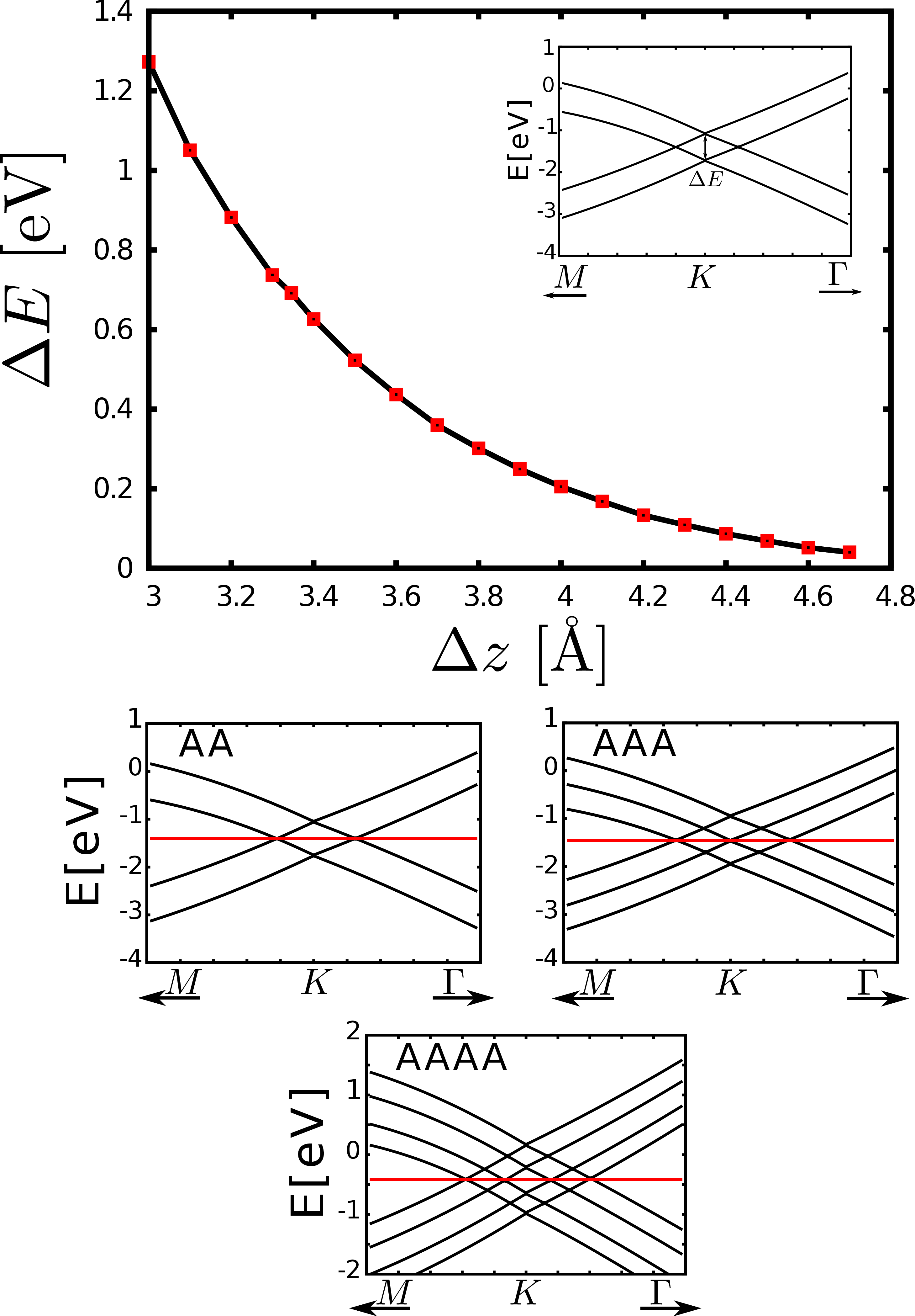}
\caption{\label{fig:AAA} (Colour online) The band structure for two, three and four freestanding 
graphene layers (bottom). Top - the dependence of $\Delta E$ shifting of the two bands of the AA 
stacking on the distance between carbon layers.}
\end{figure}

In order to determine the properties of the graphene AA stacking sequence, the bi-, three- and 
four-layer freestanding graphene structure was simulated using DFT calculations.  The resulting 
band structure is shown in Fig \ref{fig:AAA}. From these results, it follows that the band 
structure is  qualitatively similar for a various number of layers. For all cases, the 
dispersion relation is linear, with the shift arising from the overlap of the adjacent carbon 
layers. For a pair of carbon layers, the overlap of $p_z$ orbitals is changed by a change 
of the interlayer distance (see Fig. \ref{fig:AAA} - top), resulting in a different mutual shift 
of the two bands. As shown in Fig. \ref{fig:AAA} (bottom), for an odd number of graphene layers, one of 
the Dirac cone is exactly at the K point and is crossed by the Fermi level as in the isolated graphene layer. 
For an even number of layers, the intersection of the bands is located symmetrically in the vicinity 
of the K point.  

In this letter, using the HRTEM images, we have demonstrated that the epitaxial graphene, 
grown on the SiC substrate, can exist in three metastable configurations, i.e. the Bernal (ABAB), 
the rhombohedral (ABCA) and the AAAA stacking sequences. It was shown that the AAAA stacking, 
by virtue of its symmetry, has pronouncedly different electronic properties, being linear 
in the vicinity of the K point. The other two configurations have their properties different 
adopting the hyperbolic dispersion relation in the vicinity of the K point. The two structures are 
affected by the presence of the SiC substrate. In contrast to that, the AAAA stacking graphene 
preserves its linear dispersion relations in the presence of the SiC substrate. It is also
demonstrated that the change of the distance between carbon sheets amounts to a mere shift of the band, 
still preserving its linear character. Thus, the obtained results open the route to mechanically 
stable, fast electronic devices, using the Dirac dispersion relation typical for the AAAA stacking.  

This profound difference in electronic properties between AAAA and other ABCA and ABAB stacking 
was confirmed by the direct plots of electron densities. It was observed that the charge pattern 
for the AAAA stacking is similar to an isolated graphene layer. It was also shown that for ABAB 
and ABCA structures charge accumulation regions are arranged into isolated islands. 
It is expected that the honeycomb pattern is more favourable for electron traffic than the insular 
one which confirms the advantage of the AAAA graphene as a material for future fast electronic devices.

\begin{acknowledgements}
This work has been partially supported by the Polish Ministry of Science and Higher Education project 670/N-ESF-EPI/2010/0 within the EuroGRAPHENE programme of the European Science Foundation.
The authors would like to thank the Faculty of Materials Science and Engineering of Warsaw University of Technology for using the JEOL JEM 3010 transmission electron microscope.
The calculations reported in this paper were performed using the computing 
facilities of the Interdisciplinary Centre for Mathematical and Computational Modelling (ICM) of the University of Warsaw. 
The research published in this paper was supported by the Polish Ministry of Science and Higher Education under the grant no. UDA-POIG.01.03.01-14-155/09-00. 
\end{acknowledgements}
\bibliography{graph}

\end{document}